\documentclass[sigconf]{acmart}
\usepackage[english]{babel}
\usepackage[utf8]{inputenc}
\usepackage{algorithm}
\usepackage[noend]{algpseudocode}
\renewcommand\footnotetextcopyrightpermission[1]{} 
\usepackage{url}
\def\BibTeX{{\rm B\kern-.05em{\sc i\kern-.025em b}\kern-.08emT\kern-.1667em\lower.7ex\hbox{E}\kern-.125emX}}

\begin{document}

%
\title{Debian Package usage profiler for Debian based Systems}

%
\author{Bharath Honnesara Sreenivasa}
\email{bsreenivasa@umass.edu}
\affiliation{
  \institution{University of Massachusetts, Amherst}
}

\author{Ajay Rajan}
\email{arajan@umass.edu}
\affiliation{%
  \institution{University of Massachusetts, Amherst}}

%

%
\begin{abstract}
The embedded devices of today due to their CPU, RAM capabilities can run various Linux distributions but in most cases they are different from general purpose distributions as they are usually lighter and specific to the needs of that particular system. In this project, we share the problems associated in adopting a fully heavyweight Debian  based system like Ubuntu  in embedded/automotive platforms and provide solutions to optimize them to identify unused/redundant content in the system. This helps developer to reduce the hefty general purpose distribution to an application specific distribution. The solution involves collecting usage data in the system in a non-invasive manner (to avoid any drop in performance) to suggest users the redundant, unused parts of the system that can be safely removed without impacting the system functionality.
\end{abstract}

%
%


%

%

%
\maketitle

\section{Introduction}
Now-a-days embedded devices like IOT systems, hand-held phones, smart TV devices and automotive platforms have become powerful enough (with capable low-power, high performance CPU\textquotesingle s and larger memory) to run a full Linux \cite{Linux} distribution like Ubuntu \cite{ubuntu}. One typical system is ARM (64-bit) architecture powered NVIDIA Tegra \cite{tegra} \cite{oh2019hardware} SOC platform DRIVE-AGX running Ubuntu based DRIVE SDK for automotive applications.

We see that when deploying OS like Ubuntu to such automotive platforms, the considerations are very different from general purpose desktop computers. Automotive platforms solve a domain-specific problem or may be used to develop software to solve a domain-specific problem but general purpose desktop Ubuntu usage is open-ended. The ecosystem of software in the Ubuntu distribution (provided by Canonical, the company backing Ubuntu) is very well suited for general purpose desktop uses. For embedded/automotive cases, there are two problems with Ubuntu:
\begin{itemize}
 \item Ubuntu does not have suitable platform-supported images. The Ubuntu distributor expects hardware companies to start with (a very limited seed image called) Ubuntu-base and install software to user\textquotesingle s /developer\textquotesingle s needs.
  \item Users/Developers generally do not specifically know the list of software to install and end up installing a larger suite of software (generally by copying desktop configuration) which is much larger than what\textquotesingle s required for the platform\textquotesingle s use-case.
\end{itemize}
In order to solve these issues, we are introducing a tool which profiles Debian \cite{debian} \cite{claes2015historical}  packages and provides a recommendation to the user based upon a score which informs the user about the unnecessary packages, so that the user can go ahead and remove them. The tool scores all the files present in the system based upon the read and write requests and the run time of the applications using those files. Each of these files are then mapped on to their respective packages to compute a score for each package and a package with higher score is a package which is being referenced more. 
\section{Background and Motivation}
The Ubuntu installation for embedded/automotive systems \cite{raghavan2005embedded} (as described in previous section) is less than ideal, causing several unnecessary software to get into the system. This additional redundant software in the system have the following problems:
\begin{itemize}
    \item If its a daemon/service, it keeps running as a background process consuming memory and CPU
    \item In cases of debugging system issues, additional software exponentially increase number of variables to bisect and root cause the failure
    \item Unused additional software increases boot-up latency, update latency and latencies in normal workflow (if they are intertwined with user\textquotesingle s processes)
    \item Unused additional software can potentially be doors to attack vectors, leading to a potential security risk
\end{itemize}

Currently these problems are being managed by the developers deploying the system and there are two ways in which this is being done:
\begin{itemize}
    \item Get the full desktop level distribution and manually remove the packages which are not required \cite{hebner2015ubuntu}
    \item Get the minimum distribution and manually install the packages which are required
\end{itemize}
The problem with the above solutions are that it is cumbersome and requires a high level of technical knowledge and even then many a time developers underestimate or overestimate the usage of many packages. Our method directly informs the users about the packages which are being used in their distribution, so that they can take an informed decision without any expert consultation.

\section{Architecture}

\begin{figure}
  \centering
  \includegraphics[width=\linewidth]{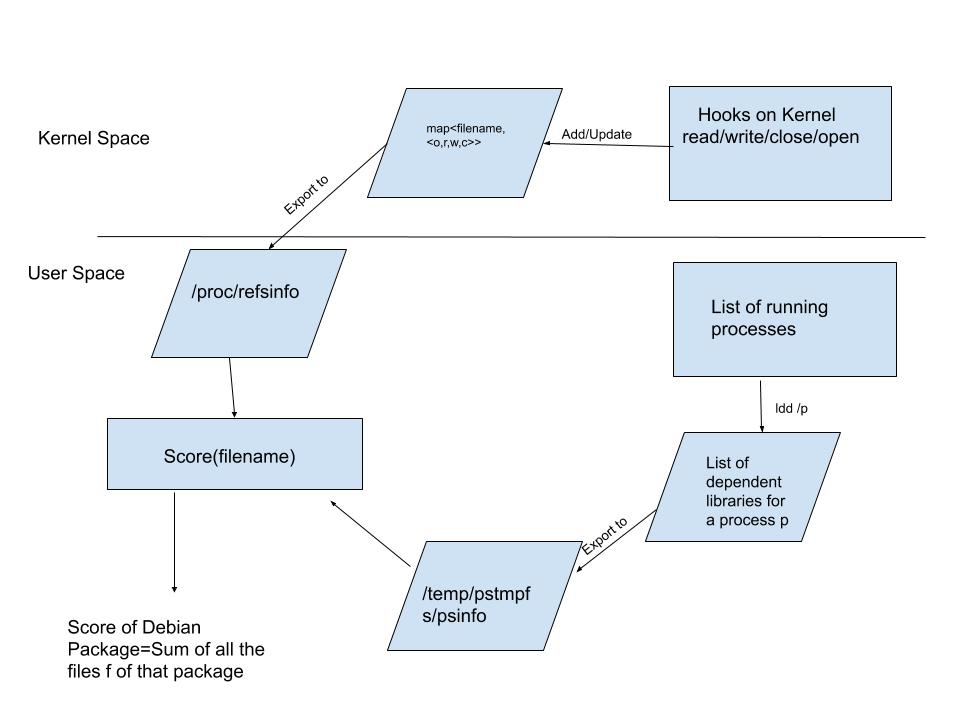}
  \caption{Architecture of Scorer, interactions between Kernel and User space}
  \Description{Architecture of Scorer}
\end{figure}

The tool is responsible for scoring the files with regards to open, close, read and write references and with regards to run time and CPU time in case they are being used by some other process and in the main memory. It consists of two parts for capturing the references: 
 \begin{itemize}
     \item For the first part, we have modified the Linux Kernel and have added a hook to the open, close and read calls to catch all the open, close and read references to the file and have stored the results into a map which has the following structure \textit{map<filename,<o,r,w,c>>} which is used by the main scorer.
     \item For the second part, we have created a shell script which updates the list of all the running processes and creates a list of all the files which are being used by the process and that list is passed on to scorer.
 \end{itemize}
The scorer gets the list from two programs and merges them using the scoring algorithm to generate a score for each file and in turn for all the packages.

\section{Design and Implementation}
The tool is used for scoring the packages and for that the tool needs to log all  references to the package files. We are using two methods which together to log all the references.
\subsection{Kernel-Space}
In the first method we are to tracking all the open,close,read and write references to the files. In order to track all the references we have added hook in the following kernel functions :
\begin{itemize}
    \item \textit{SYSCALL\_DEFINE3(open, const char \_\_user *, filename, int, flags, umode\_t, mode)} - This is responsible for opening a file given it filename. We also account for both open and openat syscalls in our profiling.
    
    \item \textit{SYSCALL\_DEFINE1(close, unsigned int, fd))} - This is responsible for closing a file given it file descriptor. We get the filename from the file descriptor as kernel keep records of opened files.
    
    \item \textit{SYSCALL\_DEFINE3(read, unsigned int, fd, char \_\_user *, buf, size\_t, count)}- This is responsible for reading a file given its file descriptor.  Similarly, we get the filename from the file descriptor as kernel keep records of opened files.
\end{itemize}

These functions are modified such that every call to these functions are recorded and logged into a file present in \textit{/proc} directory. Putting up a hook in the kernel space allows us to capture every reference to a file which would not have been possible if this was done in the user-space.

\begin{algorithm}[H]
\caption{Scorer}\label{alg:euclid}
\begin{algorithmic}[1]
\Procedure{Scorer}{}\Comment{Scorer Program}
\State $kmap<filename, <nopen, nread, nclose>>$
\State $smap<filename, <score1, score2, total>>$
\State $pmap<package, score>$

\For{Every record in \textit{/proc/refsinfo}}    
    \State $getcsv<>$
    \State $filename \gets csv[0]$
    \State $kmap[filename][0] \gets csv[1]$
    \State $kmap[filename][1] \gets csv[2]$
    \State $kmap[filename][2] \gets csv[3]$
\EndFor
\\
\For{Every k, v in kmap}   
    \State $filename \gets k$
    \State $smap[filename][0] \gets score1(v[0], v[1], v[2])$
\EndFor
\\
\For{Every record in \textit{/tmp/pstmpfs/psinfo}} 
    \State $filename, te, tc \gets csv[0], csv[1], csv[3]$  
    \State $smap[filename][1] \gets score2(te,tc)$ 
    \For{Each lpath in \textit{ldd filename}}
    \State $smap[lpath][1] \gets smap[lpath]+map[filename][1]$
    \EndFor
\EndFor
\\
\For{Every k in smap}
    \State $smap[k][2] \gets Wf * smap[k][0] + Wr*smap[k][1]$
\EndFor
\\
\State $pkglist \gets shell(dpkg list of installed packages in the system)$
\For{Every pkg in \textit{$pkglist$}}
    \State $pmap[pkg] \gets 0$
\EndFor
\\
\For{Every path in $smap$}
    \State $pkg \gets shell(dpkg path)$
    \State $pmap[pkg] \gets pmap[pkg] + smap[path][2]$
\EndFor
\\
\State \textbf{return} $pmap$
\EndProcedure
\end{algorithmic}
\end{algorithm}

The following is the implementation in kernel-space
\begin{itemize}
    \item Add a function to each of read,write, open and close hooks to count on filename
    \item Add a /proc/refsinfo entry into kernel procfs.
    \item Keep count of number of open, close and read operations for each filename in a hashmap.
    \item Implement /proc read driver to expose hashmap data in csv format.
\end{itemize}

\begin{algorithm}[H]
\caption{Kernel Filesystem Profiling and Exporting to Userspace}\label{alg:euclid}
\begin{algorithmic}[2]
\Procedure{fs-profiling}{}\Comment{Profiling filesystem}
\State $kmap<filename, <nopen, nread, nclose>>$

\For{Every open to $filename$}
    \State Lookup filename in $kmap$
    \If {filename does not exist}
        \State $kmap[filename] \gets <0, 0, 0>$
    \EndIf
    \State $kmap[filename][0] \gets kmap[filename][0] + 1$
\EndFor
\\
\For{Every read to $fd$}
    \State Lookup filename from fd to map to $kmap$
    \State $kmap[filename][1] \gets kmap[filename][1] + 1$
\EndFor
\\
\For{Every close to $filename$}
    \State Lookup filename from fd to map to $kmap$
    \State $kmap[filename][2] \gets kmap[filename][2] + 1$
\EndFor
\EndProcedure
\\
\Procedure{read /proc/refsinfo}{}
\For{Every k, v in $kmap$}
    \State Convert into csv string "k,v[0],v[1],v[2]"
    \State output csv string
    \State output newline
\EndFor
\EndProcedure
\end{algorithmic}
\end{algorithm}

\subsection{User-Space Profiling}
There are scenarios where the files are copied on to primary memory while booting the system up, in order efficiently track those files we are monitoring the processes which are running in the system.
In order to monitor all processes we are using a shell script which checks the currently running process every 1 second and updates the a log file which is present in \textit{/tmpfs} directory.

The \textbf{Scorer} takes the two files i.e. the file generated by the kernel and the file generated by the shell script.For every process in the list we get all the libraries being used by that particular process and maps those files in the result obtained from the first file. So now we have a structure with the information like read/open/close references, CPU time and execution time. We use all this information to compute a score for each file.

\subsubsection{Scorer}
Scorer consists of algorithms which are used to score every Debian Package and that score is used to profile the packages. A score for a given Debian Package is the sum of scores of all the files contained in the package.
\begin{equation}
S_{File}= S_{FS} + S_{PS}
\end{equation}
$FS=$ Number I/O references to that file
\\ PS= Running time of the process using that file

\begin{equation}
\small
S_{FS}= B(N_{Open}-N_{Close}) + W_{Open}(N_{Open}) + \\ W_{Read}(N_{Read})
\end{equation}
\\$B=$ Open Bonus \\
$W\textsubscript{Open},W\textsubscript{Read},W\textsubscript{Close} =$ Weight for open, read and close \\ 
$N\textsubscript{Open}, N\textsubscript{Read}, N\textsubscript{Close} =$ Number of open, read and close to the file \\ \\
For S\textsubscript{FS} we are considering number of opens, reads and close to a file and on top of that we have provided a bonus of 100 points for the files which are still in open state i.e if the file is not closed it is being actively used by some process or system. Default weights for open and read operations are 1 and 5 respectively but they can be configured by the user as per the needs.

\begin{equation}
S_{PS}= W_{Elapsed}(Time_{Elapsed}) + W_{CPU}(Time_{CPU})    
\end{equation}
$W\textsubscript{Elapsed}, W\textsubscript{CPU}=$ Weight for Elapsed, CPU Time \\
$Time\textsubscript{Elapsed}, Time\textsubscript{CPU}=$ Elapsed and CPU time for a Process
\\ 
For S\textsubscript{PS} we are considering the running and CPU times of the process that are using the given file.The Weight given to the CPU time is higher than the weight given to the elapsed time because there can be multiple processes in the primary memory which are just in running state but are not utilizing the CPU.  

\section{Experiments}
The kernel-space and user=space components are executed together and data gather on the system w.r.t usage of various files and Debian packages in the system.

\subsection{Setup}
There two parts of scoring filesystem references are read from /proc/refsinfo and executables and daemons run on the system are accounted through a users-space profiling module written in bash shell scripting which executes ps to return elapsed time since application/daemon has started and the actual CPU time used. We have profiling of filesystem references and processes profiled in our setup: A x86-64 Intel(R) Core(TM) i7-6800K CPU running at 3.40GHz, memory of 32GB, 1TB of SSD and running Ubuntu 16.04 (64-bit). The scorer application is executed to obtain scored metrics for system usage. The system has been running since 85 days.

\begin{figure}
  \centering
  \includegraphics[width=\linewidth]{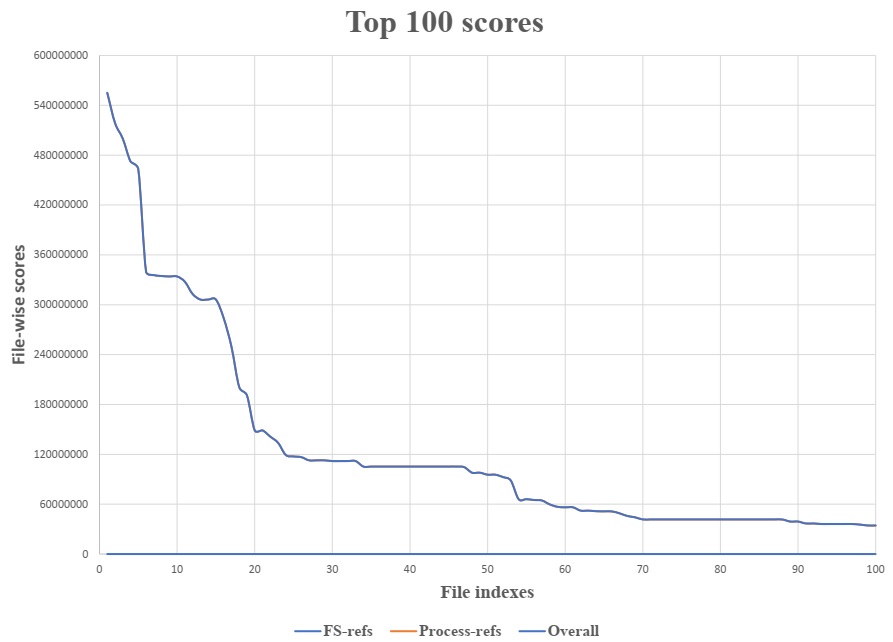}
  \caption{File-wise scores over the system generated by Scorer}
  \Description{File-wise scores over the system generated by Scorer}
\end{figure}

\begin{figure}
  \centering
  \includegraphics[width=\linewidth]{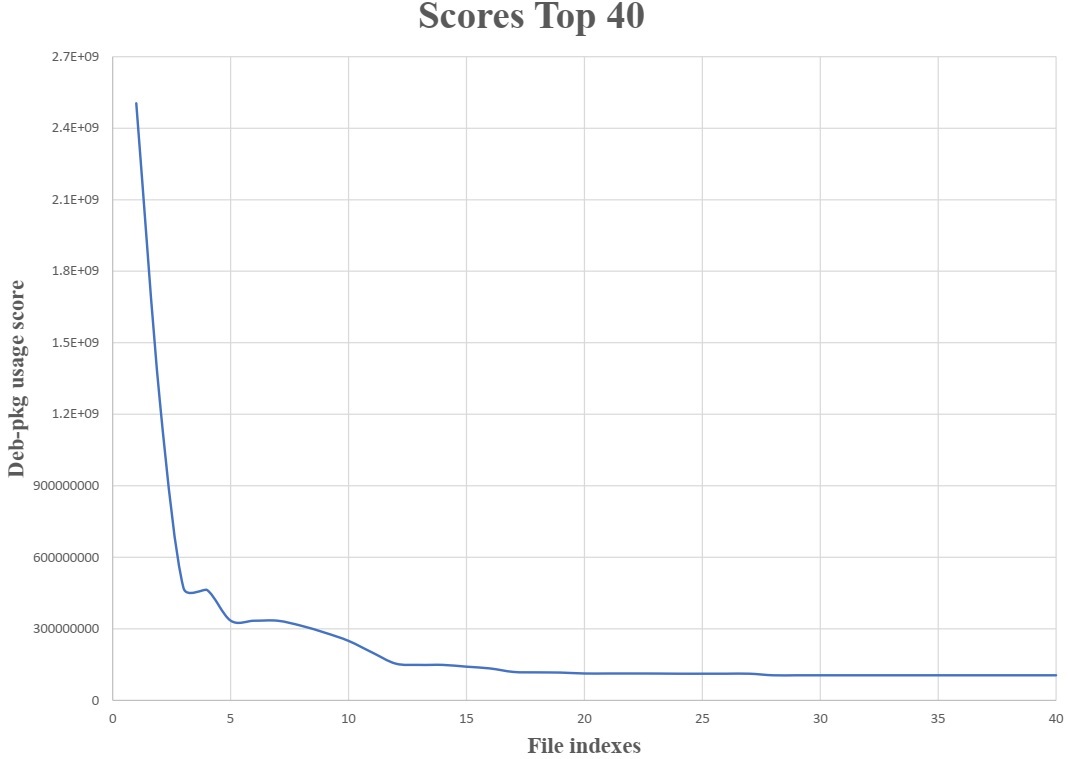}
  \caption{Debian Package-wise scores over the system generated by Scorer}
  \Description{Debian Package-wise scores over the system generated by Scorer}
\end{figure}

\begin{figure}
  \centering
  \includegraphics[width=\linewidth]{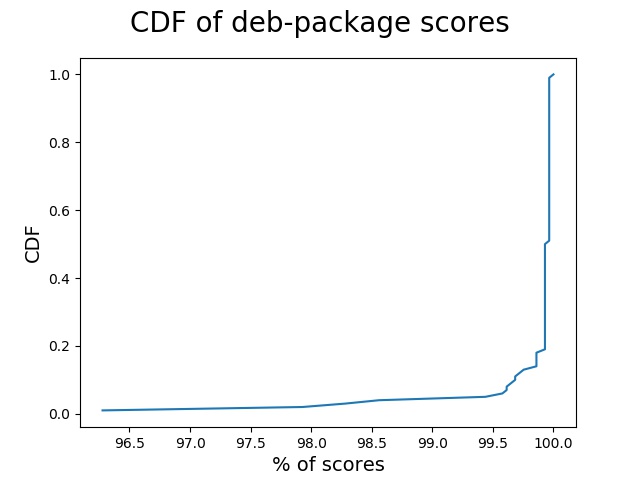}
  \caption{CDF of debian package scores}
  \Description{CDF of debian package scores}
\end{figure}

\begin{figure}
  \centering
  \includegraphics[width=\linewidth]{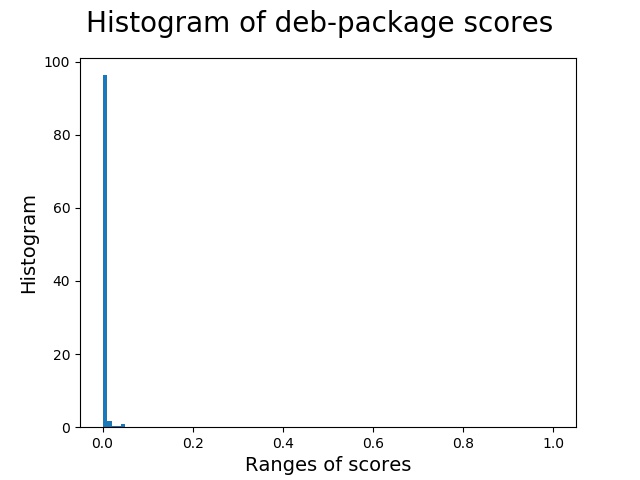}
  \caption{Histogram of scores for debian packages}
  \Description{Histogram of scores for debian packages}
\end{figure}

\begin{figure}
  \centering
  \includegraphics[width=\linewidth]{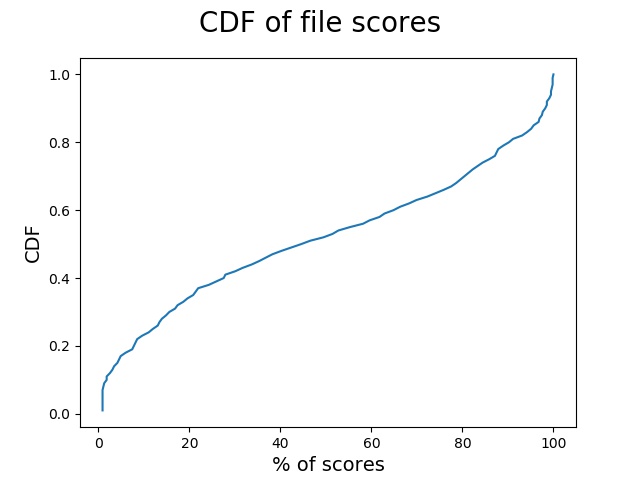}
  \caption{CDF of file-wise scores}
  \Description{CDF of file-wise scores}
\end{figure}

\begin{figure}
  \centering
  \includegraphics[width=\linewidth]{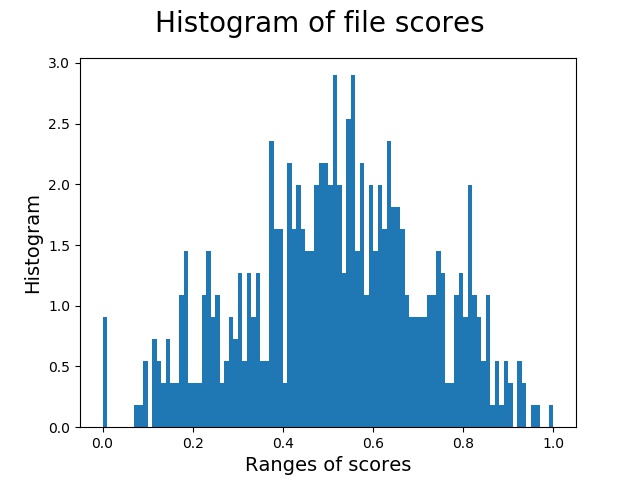}
  \caption{Histogram of file-wise scores}
  \Description{Histogram of file-wise scores}
\end{figure}

\subsection{Observations}
From the list executables and libraries set we see that overall libraries show much more usage and executables. We see top 3 libraries scoring (M - Million, B - Billion) around 500 M are:
\begin{itemize}
    \item \text{/lib/x86\_64-linux-gnu/libc.so.6}
    \item \text{/lib/x86\_64-linux-gnu/libdl.so.2}
    \item \text{/lib/x86\_64-linux-gnu/libpthread.so.0}
\end{itemize}
The top 3 executables scoring around 7 M are:
\begin{itemize}
    \item \text{/usr/lib/xorg/Xorg}
    \item \text{/usr/bin/dockerd}
    \item \text{/usr/bin/lxpanel}
\end{itemize}
The decreasing graph trend shows scores decreasing from maximum to one-fifth of maximum value over 20 files in the system.
\\\\
The scores of debian packages show that top 3 scores for packages are:
\begin{itemize}
    \item \text{libc6 with score 2 B}
    \item \text{libglib2.0-0 with score 1 B}
    \item \text{libpcre3 with score 500 M}
\end{itemize}
The graph trend for packages show score falls to one-hundredth of maximum within 5 packages.

\section{Future Work}
Based on limitations and observations following future improvements can be proposed:
\begin{itemize}
\item Model usage of kernel modules using ftrace account back to the current weighted model
\item Access to dpkg database for every-file takes lot of I/O operations, this can be optimized
\item The current kernel implements bottlenecks any open, read or close to any filesystem in by contending access to hashmap, this should be optimized to keep overheads to a minimum
\item Analysis to be done to better relate weights for filesystem references, elapsed time and CPU time. This is ensure more uniform scores over the system.
\end{itemize}

\section{Conclusions}
As seen from the data, we have sharp peaks of usage for set of files or set of debian packages which goes down rapidly to small scores. This indicates only a core set of files are extremely important for the system. We also see large number of installed packages with zero scores which show they are not in active use in the system they may have been installed inadvertently by the user or the distribution default configuration.

\bibliographystyle{ACM-Reference-Format}
\bibliography{sample-base}


\end{document}